\documentclass[aps, reprint, superscriptaddress, amsmath, amssymb]{revtex4-2}
\usepackage[utf8]{inputenc}
\usepackage[T1]{fontenc}
\usepackage{newtxtext,newtxmath} 

\usepackage{graphicx}   
\usepackage{dcolumn}    
\usepackage{bm}         
\usepackage{tabularx}
\usepackage{siunitx}
\usepackage{textcomp}
\graphicspath{{Figures/}}

\usepackage{xcolor}
\usepackage{color, soul} 
\usepackage[normalem]{ulem} 
\usepackage{comment}

\newcommand{\ba}{\begin{eqnarray}}
\newcommand{\ea}{\end{eqnarray}}

\begin{document}
\title{Measuring Reactive-Load Impedance with Transmission-Line Resonators Beyond the Perturbative Limit}

\author{Xuanjing Chu}
\altaffiliation{These authors contributed equally to this work.}
\affiliation{Department of Applied Physics and Applied Mathematics, Columbia University, New York, NY 10027, USA}
\author{Jinho Park}
\altaffiliation{These authors contributed equally to this work.}
\affiliation{Department of Mechanical Engineering, Columbia University, New York, NY 10027, USA}
\author{Jesse Balgley}
\affiliation{Department of Mechanical Engineering, Columbia University, New York, NY 10027, USA}
\author{Sean Clemons}
\affiliation{Department of Computer Science, Howard University, Washington, DC 20059, USA}
\author{Ted S. Chung}
\affiliation{Department of Mechanical Engineering, Columbia University, New York, NY 10027, USA}
\author{Kenji Watanabe}
\affiliation{National Institute for Materials Science, 1-2-1 Sengen, Tsukuba, Ibaraki 305-0044, Japan}
\author{Takashi Taniguchi}
\affiliation{National Institute for Materials Science, 1-2-1 Sengen, Tsukuba, Ibaraki 305-0044, Japan}
\author{Leonardo Ranzani}
\affiliation{RTX BBN Technologies, Quantum, Photonics, and Computing Group, Cambridge, MA 02138, USA}
\author{Martin V.~Gustafsson}
\affiliation{RTX BBN Technologies, Quantum, Photonics, and Computing Group, Cambridge, MA 02138, USA}
\author{Kin Chung Fong}%
\affiliation{Department of Physics, Northeastern University, Boston, MA, USA}
\affiliation{Quantum Materials and Sensing Institute, Northeastern University, Burlington, MA, USA}
\affiliation{Department of Electrical and Computer Engineering, Northeastern University, Boston, MA, USA}
\author{James Hone}
\email{jh2228@columbia.edu}
\affiliation{Department of Mechanical Engineering, Columbia University, New York, NY 10027, USA}
\date{\today}

\begin{abstract}
We develop an analytic framework to extract circuit parameters and loss tangent from superconducting transmission-line resonators terminated by reactive loads, extending analysis beyond the perturbative regime. The formulation yields closed-form relations between resonant frequency, participation ratio, and internal quality factor, removing the need for full-wave simulations. We validate the framework through circuit simulations, finite-element modeling, and experimental measurements of van der Waals parallel-plate capacitors, using it to extract the dielectric constant and loss tangent of hexagonal boron nitride. Statistical analysis across multiple reference resonators, together with multimode self-calibration, demonstrates consistent and reproducible extraction of both capacitance and loss tangent in close agreement with literature values. In addition to parameter extraction, the analytic relations provide practical design guidelines for maximizing energy participation ratio in the load and improving the precision of resonator-based material metrology.

\end{abstract}

\maketitle

\section{Introduction}
Accurate microwave-impedance measurements are essential for probing electronic properties and evaluating material performance in quantum circuits. While capacitance and inductance constitute the imaginary components of the complex impedance, careful analysis of these reactive elements can reveal key insights into electronic structure and correlated states through the quantum capacitance~\cite{Xia2009, Riley2015, Maji2024} and kinetic inductance~\cite{Kreidel2024, Banerjee2025}. In contrast, the resistive component captures energy dissipation within the material, \textit{i.e.}, the dielectric loss in a capacitor under the equivalent series resistance model. This dielectric loss is a primary factor limiting coherence times in superconducting qubits~\cite{Larsen2015, Lange2015, casparis2018, Wang2022, strickland2024, zheng2024}, and accurate quantification of such loss is therefore essential for improving qubit performance~\cite{Wang2015}. Moreover, minimizing dielectric loss also benefits other superconducting devices, including amplifiers, detectors, and parametric circuits, thereby advancing the full suite of superconducting quantum technologies.


Resonant circuits provide a highly sensitive means of measuring the complex impedance of materials or circuit components. When the device under test is small --- such as a lumped element in a quantum circuit or a two-dimensional material with exotic electronic properties --- it can be incorporated into a resonator as a load that produces measurable shifts in the resonant frequency $f_r$ and internal quality factor $Q_i$~\cite{Kreidel2024}. To put this sensitivity in perspective, attaching a $200~\mathrm{fF}$ capacitor with an equivalent series resistance of $200~\mathrm{\mu\Omega}$ to a resonator with $f_r = 7~\mathrm{GHz}$ and $Q_i = 10^6$ shifts $f_r$ by approximately $830~\mathrm{MHz}$ and reduces $Q_i$ to $8\times10^5$. The sensitivity of this approach improves further with increasing $Q_i$ and greater electromagnetic participation of the load.

Three-dimensional cavities can achieve $Q_i > 10^9$ and support multiple modes~\cite{Reagor2013, Lei2023, Checchin2021}, but their macroscopic mode volumes limit coupling to nanoscale samples. Planar superconducting resonators overcome this limitation by confining electromagnetic fields near the surface, enabling strong coupling to nanoscale devices while retaining straightforward fabrication and modeling~\cite{Krantz2019, Blais2021}. Lumped-element resonators concentrate fields into small volumes and thereby maximize coupling to nanoscale devices, though their nonuniform field distribution complicates loss modeling and parameter extraction. In contrast, transmission-line resonators --- typically realized on-chip as coplanar waveguides (CPWs) --- combine ease of fabrication, characterization, and modeling with the ability to support multiple harmonic modes~\cite{Jin2025,Megrant2012, Kalacheva2023, Antony2021}. However, most analyses of transmission-line resonators rely on perturbative approximations~\cite{Gao2008, Kreidel2024, McRae2020}, which can limit accuracy, or on numerical simulations~\cite{Wang2022, Wenner2011}, which demand specialized expertise and significant computational resources for evaluating participation ratios.

In this work, we derive closed-form analytic expressions linking the resonant frequency, participation ratio, and internal quality factor of a reactive load coupled to a transmission-line resonator. Whereas previous microwave analyses rely on perturbative approximations, the present formulation provides a unified treatment valid for arbitrary load reactances, removing the need for full-wave electromagnetic simulations. Our analytic results, particularly Eqs.~\ref{eq:resofreq}, \ref{eq:p}, and \ref{eq:Qi_p}, identify resonator configurations that maximize the accuracy of complex impedance extraction and provide a framework for improved device characterization through comparison of responses across multiple resonant modes. In particular, we highlight an operating point at $\lvert X\rvert \approx Z_0$, which maximizes energy-participation ratio of the reactive load --- henceforth referred to as the device under test (DUT) --- and thus enhances parameter sensitivity while reducing extraction uncertainty. This condition of maximization of energy-participation ratio in the DUT and measurement sensitivity can be compared with the condition of conjugate matching that maximizes power transfer into a load. This framework enables systematic optimization of microwave characterization for low-dimensional materials and quantum devices. Validation through circuit simulations, finite-element modeling, and experimental measurements of resonators terminated by van der Waals capacitors confirms excellent agreement between theory and experiment.

\begin{figure}[t]
    \centering
    \includegraphics[width=0.8\columnwidth]{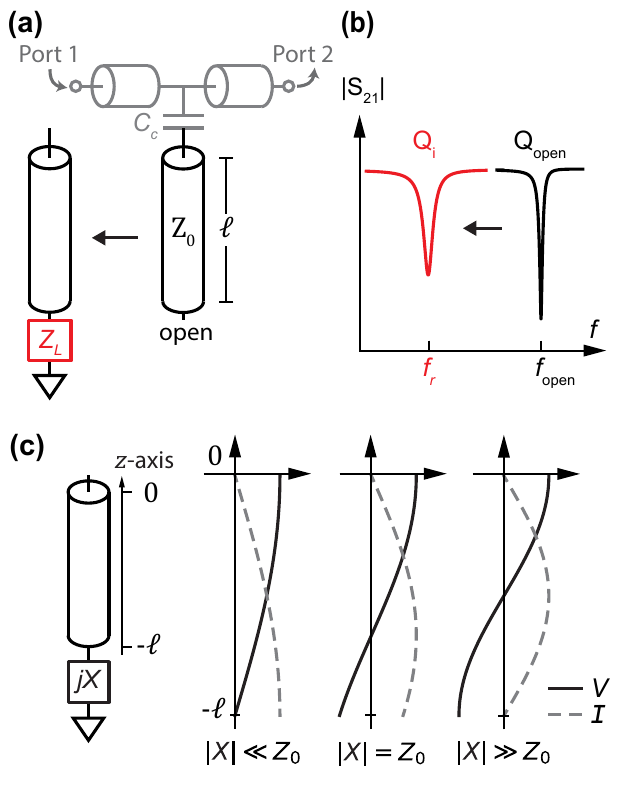}
    \caption{
    (a) Schematic of a waveguide resonator with characteristic impedance $Z_0$ and length $\ell$, terminated by a DUT characterized by $Z_L = r + jX$. The resonator is coupled to the busline through a coupling capacitor $C_c$.
    (b) The reactive termination shifts the resonant frequency from $f_\text{open}$  to $f_r$, and changes the internal quality factor from $Q_\text{open}$ to $Q_i$. The two corresponding ports for $S$-parameter measurement are labeled in (a). 
    (c) Standing wave voltage (solid) and current (dashed) profiles along the resonator length for three representative reactive terminations: $\lvert X\rvert  \ll Z_0$, $\lvert X\rvert  = Z_0$, and $\lvert X\rvert  \gg Z_0$. The termination is modeled as a purely reactive load ($jX$), shown in the schematic. The plots illustrate the capacitive terminations, as current lags the voltage by $\pi/2$ at the DUT.
    }
    \label{fig1}
\end{figure}

\section{Theoretical Model} 
The theoretical framework is illustrated in Fig.~\ref{fig1}(a). A waveguide with length $\ell$ and characteristic impedance $Z_0$ creates a notch-type transmission-line resonator with resonant frequency $f_\text{open}$, and internal quality factor $Q_\text{open}$. The resonator is capacitively coupled to a busline, and measurement of the transmission $S_{21}$ provides readout of the resonator properties (Fig.~\ref{fig1}(b)). While we focus on capacitive coupling in the present discussion, the framework can be readily extended to inductive coupling. The DUT has impedance $Z_L = r + jX$, with $r$ and $X$ denoting resistance and reactance, respectively. For a capacitor, the reactance is $X=-1/(2\pi fC)$, while for an inductor it is $X=2\pi fL$. When the resonator is terminated by the DUT, its resonant frequency shifts to $f_r$ and the quality factor changes to  $Q_i$. As will be detailed below, the frequency shift can be used to measure $X$, and the change in quality factor can be used to measure dissipation as quantified by the loss tangent $\tan\delta=r/\lvert X\rvert $.

The analysis below considers the regime where dissipative contributions are small compared to reactive ones, both in the resonator and in the DUT ($\tan\delta \ll 1$). This matches experimental conditions for planar superconducting resonators, which typically exhibit quality factors in the range of $10^5$–$10^7$, and for DUTs consisting of low-loss capacitors or inductors. 



Examples of the standing-wave voltage $(V)$ and current $(I)$ profiles in a transmission-line resonator capacitively coupled to a busline are shown in Fig.~\ref{fig1}(c). When the coupling capacitance $C_c$ is sufficiently small, the resonator exhibits a voltage antinode (and corresponding current node) at one end ($z=0$). The boundary condition is then determined by the termination at the opposite end ($z=-\ell$). For $\lvert X\rvert  \ll Z_0$ (\textit{i.e.} small inductance or large capacitance), a voltage node forms at the DUT, giving rise to a fundamental $\lambda/4$ resonant mode. Conversely, when $\lvert X\rvert  \gg Z_0$ (small capacitance or large inductance), a current node forms at the DUT, producing a $\lambda/2$ mode. These limits define perturbative regimes, in which the DUT participation can be treated as a small inductive or capacitive perturbation to an unloaded $\lambda/4$ or $\lambda/2$ resonator, respectively \cite{McRae2020, Kreidel2024, Antony2021}.

In contrast, when the DUT reactance is comparable to the characteristic impedance of the resonator ($\lvert X\rvert  \sim Z_0$), the end is neither a voltage nor a current node, and a significant fraction of the electromagnetic energy in the system is stored in the DUT. In this regime, the resonance becomes highly sensitive to the loss in DUT.



\section{Analytic solution and simulation}

\begin{figure}[t]
    \centering
    \includegraphics[width=0.76\columnwidth]{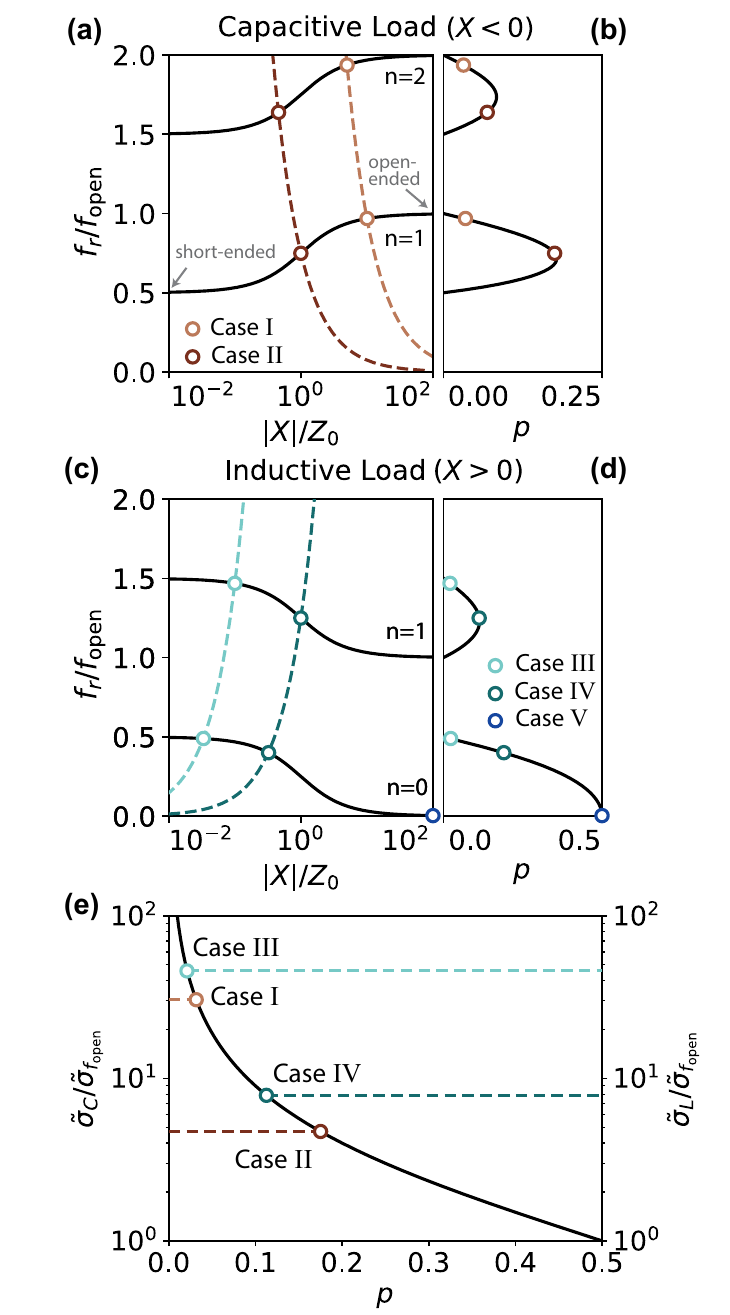}
    \caption{
    (a–b) Normalized resonant frequencies $f_r/f_\text{open}$ (a) and corresponding DUT participation ratios $p$ (b) as functions of normalized capacitive reactance $\lvert X\rvert /Z_0$ ($X<0$) for modes $n=1$ and $n=2$. Here, $Z_0=50~\Omega$. The short-ended ($\lambda/4$) and open-ended ($\lambda/2$) limits are indicated for reference.
    (c–d) Corresponding results for inductive reactance ($X>0$), shown for modes $n=0$ and $n=1$.
   In panels (a–d), dashed curves show the frequency-dependent reactance of example DUT (capacitors in brown, inductors in teal). For a 7 GHz resonator, the capacitors are chosen as 47 fF (Case I) and 606 fF (Case II), and the inductors as 77 pH (Case III) and 909 pH (Case IV). At resonance, Cases I and III correspond to $X = -500~\Omega$ and $X = +5~\Omega$, representing the perturbative regime. In contrast, Cases II and IV yield $\lvert X\rvert \approx 50~\Omega$, corresponding to conditions of maximal DUT energy participation ratio. The resonant frequency for each DUT is obtained from the intersection of its dashed reactance curve with the solid resonance curve, indicated by open circles. The corresponding participation ratios are shown in (b) and (d).
    (e) Relative uncertainty in extracted capacitance $\tilde\sigma_C = \sigma_C/C$ and inductance $\tilde\sigma_L = \sigma_L/L$ as a function of DUT energy participation ratio $p$. Uncertainties are normalized to the baseline frequency uncertainty of a reference resonator, $\tilde\sigma_{f_\text{open}} = \sigma_{f_\text{open}}/f_\text{open}$. Results are shown for Cases I–IV, defined consistently across panels (a)–(d).
    }
    \label{fig2}
\end{figure}

To calculate the resonant frequency, we solve the transmission-line equations~\cite{Pozar} using the appropriate boundary conditions (see Appendix~\ref{appendix:f_derivation}) and obtain
\begin{align}
f_r = \left(\frac{1}{\pi}\tan^{-1}\frac{Z_0}{X}+n\right)f_\text{open}, \label{eq:resofreq}
\end{align}
where $n$ is the standing-wave mode index. This result, consistent with earlier analyses in Ref.~\cite{Ramachandran1961}, provides a concise expression for the resonance of a transmission-line resonator terminated by an arbitrary reactive load. Although simple in form, it serves as the foundation for thoroughly extracting circuit parameters when extended to include higher-order modes, dissipation, and the concept of the energy participation ratio, as developed in this section.

Figs.~\ref{fig2}(a) and (b) plot the normalized resonance frequency $f_r/f_\text{open}$ (solid black lines) as a function of $\lvert X\rvert /Z_0$ for capacitive and inductive loads, respectively. As the reactance increases, the frequencies of each resonant mode $n$ increase (decrease) in the capacitive (inductive) case.

For a given capacitive or inductive DUT, the frequency-dependent value of $X$ determines the value of $f_r$.  As illustrative examples, Fig.~\ref{fig2}(a) shows two capacitive terminations (47 fF, Case I; and 606 fF, Case II), and Fig.~\ref{fig2}(c) shows two inductive terminations (77 pH, Case III; and 909 pH, Case IV). The dashed curves display the corresponding reactances, evaluated assuming $f_\text{open} = 7$ GHz. The resonant condition in each case is given by the intersection of the solid and dashed curves (open circles). At resonance, Cases I and III correspond to $X=-500~\Omega$ and $X=+5~\Omega$, placing the system in the perturbative regime. Cases II and IV correspond to $\lvert X\rvert  = Z_0 = 50~\Omega$, where the resonator operates near the point of maximal DUT energy participation ratio.

The DUT participation ratio, which quantifies the fraction of electromagnetic energy stored in the DUT, dictates the accuracy to which DUT properties can be measured. By integrating the electrical and magnetic energy profiles in the loaded resonator (see Appendix \ref{appendix:p_derivation}), we obtain:
\begin{align}
p &= \frac{\lvert \sin \phi \rvert}{\phi + \lvert \sin \phi \rvert},\label{eq:p} 
\end{align}
where $\phi = 2\pi f_r / f_\text{open}$ quantifies the phase shift between the incident and reflected wave due to the DUT.

Figs.~\ref{fig2}(b) and (d) show $p$ as a function of $f_r/f_\text{open}$ for capacitive and inductive terminations, respectively. For each, $p$ is bounded by the envelope $1/(1+f_r/f_\text{open})$, yielding a general reduction in maximum participation for higher harmonics. 

As shown in Fig.~\ref{fig2}(b) and (d), the DUT participation $p$ exhibits a local maximum on the $n=1$ branch when the reactance satisfies $\lvert X\rvert  = Z_0$ (Cases II and IV), corresponding to a regime in which a substantial fraction of the electromagnetic energy is stored in the DUT. This behavior follows directly from Eq.~\ref{eq:p} and persists for higher-order modes, as visualized in Fig.~\ref{fig2}(b). A special situation arises for the $n=0$ branch. For capacitive terminations, the $n=0$ mode would require a negative resonant frequency and is therefore unphysical. For inductive terminations, however, the $n=0$ mode is allowed and $p$ can in principle approach 50\% at large inductance (Case V). In practice, the resonant frequency of this mode approaches zero, rendering it experimentally inaccessible. The more apt choice is then to design the reactance to still have significant participation but not shift the resonance out of the practically measurable bandwidth, as exemplified in Case IV. 





We next examine how variations in $f_r$ and $f_\text{open}$ affect the accuracy of the extracted reactance. For the DUTs studied here, $f_r$ can be determined with part-per-million accuracy, such that its uncertainty is negligible. In contrast, $f_\text{open}$ cannot be determined directly for the loaded resonator. Instead, $f_\text{open}$ must by estimated by measuring a nominally identical open resonator, and is thus affected by variability of the fabrication process. By measuring more than 50 reference resonators, we find a relative variation in $f_{\text{open}}$ of $\tilde{\sigma}_{f_{\text{open}}} = 0.14\%$, where $\tilde{\sigma}_{f_{\text{open}}} = \sigma_{f_{\text{open}}}/f_{\text{open}}$ denotes the fractional uncertainty. Similarly,  $\tilde{\sigma}_L=\sigma_L/L$ and $\tilde{\sigma}_C=\sigma_C/C$. The relative uncertainty in the extracted capacitance (or inductance) takes the form:
\begin{align}
\tilde \sigma_{L,C}
\approx \frac{1 - p}{p} 
\tilde\sigma_{f_\text{open}}.
\end{align}

This expression confirms the crucial role played by $p$, since uncertainty is minimized when $p$ is maximized. To visualize this dependence, Fig.~\ref{fig2}(e) shows $\tilde\sigma_C$ and $\tilde\sigma_L$, normalized to $\tilde\sigma_{f_\text{open}}$, as a function of $p$. DUTs with higher participation (Case II and IV) systematically yield nearly an order of magnitude lower uncertainty than low-participation DUTs (Case I and III) under the same baseline frequency variability, thereby providing superior accuracy for measuring capacitance or inductance. The max DUT energy participation ratio mode can be realized either by modifying the DUT design or by tuning the resonator frequency.


The need to estimate $f_{\text{open}}$ can be removed by measuring multiple harmonics of the same loaded resonator. From Eq.~\ref{eq:resofreq}, the unknown parameters are $f_{\text{open}}$ and the DUT capacitance or inductance. Thus, measurements of two resonant modes provide two equations for these two unknowns, enabling numerical solution without reference to a separate resonator. We demonstrate this multimode self-calibration procedure in Sec.~IV.

Access to additional modes ($n>2$) further overdetermines the system, allowing validation of the extracted parameters. Moreover, because capacitive and inductive reactances have opposite signs and distinct frequency dependences, comparing $X$ across modes separates their respective contributions to the total reactance. This, in turn, enables the identification of parasitic effects, such as stray capacitance or geometric inductance.

In addition to shifting the resonant frequency, the DUT termination also modifies the internal quality factor of the resonator. By comparing the internal quality factors of DUT-terminated resonator ($Q_i$) and a nominally identical open resonator ($Q_{\text{open}}$), one can, in principle, extract the loss tangent $\tan\delta$ (full derivation in Appendices~\ref{appendix:p_derivation} and \ref{appendix:q_derivation}). The resulting expression for the internal quality factor is
\begin{align}
Q_i^{-1} &= \frac{2\lvert \sin \phi \rvert \tan \delta + 2 \pi Q^{-1}_\text{open}}{\phi + \lvert \sin \phi \rvert}, \\
\tan\delta&= \frac{1}{2p}Q_i^{-1}-\frac{1-p}{2p}\frac{2\pi}{\phi}Q^{-1}_\text{open}.\label{eq:Qi_p}
\end{align}
Here, $Q_{\text{open}}=\pi/(2\alpha l)$, $\alpha$ is the attenuation constant per unit length for the transmission-line resonator. In the limit $\phi \rightarrow n\pi$, these analytic results reduce to the perturbative forms obtained in previous treatments \cite{McRae2020, Kreidel2024, Antony2021}. However, for arbitrary $\phi$, the system can deviate substantially from the perturbative regime, and Eq.~\ref{eq:Qi_p} provides an explicit analytic relation between the resonator loss and reactive loading that was not available in previous microwave analyses. Together, Eqs.~\ref{eq:resofreq}, \ref{eq:p}, and \ref{eq:Qi_p} form a unified formulation valid beyond the perturbative regime, enabling quantitative extraction of reactive and dissipative parameters.


\begin{figure}[t]
\centering
\includegraphics[width=1\columnwidth]{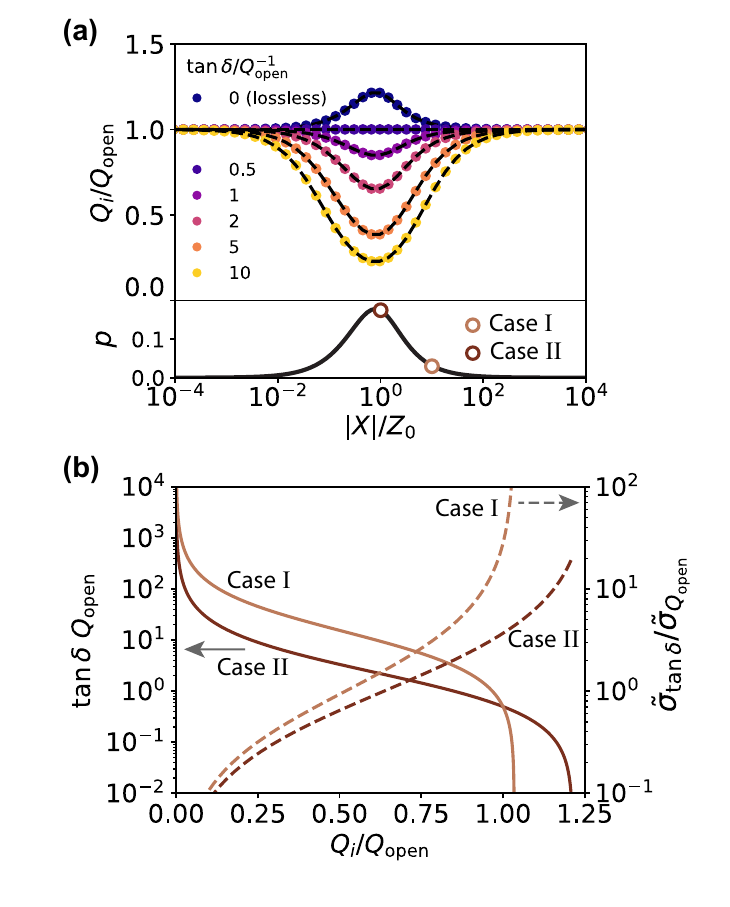}
\caption{
    (a) Simulated results (circles) for capacitive DUT at fundamental mode $n=1$ obtained using \textit{Qucs} for selected values of the loss tangent closely match the analytic predictions (dashed lines), computed by substituting the same loss tangent values into the model. 
    Lower panel: DUT energy participation ratio $p$ versus $\lvert X\rvert /Z_0$ for a capacitive DUT configuration. The maximum achievable participation in this setup is approximately $17.85\%$. The two open circles reproduce the capacitive Cases I and II from Fig.~\ref{fig2}(a).
    (b) Extracted DUT loss tangent for the two representative Cases in (a). Solid lines show $\tan\delta~Q_\text{open}$ as a function of $Q_i/Q_\text{open}$ (left axis), and dashed lines show the corresponding relative uncertainties $\sigma_{\tan\delta}/\tan\delta$, normalized to the baseline quality factor uncertainty of a reference resonator (right axis).
    }
\label{fig3}
\end{figure}


To test the analytic model, we simulate a CPW resonator with $Z_0 = 50~\Omega$ terminated by a series combination of a resistor $r$ and capacitor $C$ using the \textit{Qucs} circuit simulation software. We set $\alpha$ proportional to $f_r$, as the loss is dominated by dielectric loss. The resonance parameters $f_r$ and $Q_i$ are extracted from the simulated spectra by circle-fit analysis of the $S$-parameters \cite{Probst_circle_fit}.

Fig.~\ref{fig3}(a) shows the normalized internal quality factor, $Q_i/Q_\text{open}$, as a function of $\lvert X\rvert /Z_0$, with values of the DUT loss tangent annotated. As expected, large DUT losses reduce $Q_i$. Conversely, DUTs with sufficiently low loss can increase $Q_i$ values \emph{exceeding} $Q_\mathrm{open}$, as a larger fraction of the electromagnetic energy is stored in the low-loss DUT rather than in the lossy transmission-line section. Analytic predictions from our formalism (dashed lines) are in excellent agreement with the circuit simulations.

In principle, it is straightforward to determine $\tan\delta$ by measuring $f_r$, $f_\text{open}$, $Q_i$, and $Q_\text{open}$, and applying Eqs.~\ref{eq:p} and \ref{eq:Qi_p}. However, statistical variation inherent to resonators limits the accuracy of such extraction. In particular, $Q_\text{open}$ exhibits variability not only between nominally identical resonators but also across repeated measurements of the same device \cite{McRae2025}. When uncertainty in $Q_\text{open}$ dominants, the fractional uncertainty in $\tan\delta$ is given by:
\begin{align}
      \tilde\sigma_{\tan \delta} \approx \frac{(1-p)}{Q_\text{open}/Q_i - (1-p)}\tilde\sigma_{Q_\text{open}}.\label{eq:sigma_tand_approx}
\end{align}
In which, $\tilde{\sigma}_{Q_\text{open}}=\sigma_{Q_\text{open}}/Q_\text{open}$ and $\tilde{\sigma}_{\text{tan}\delta}=\sigma_{\text{tan}\delta}/\tan{\delta}$. 
For a given variability in $Q_{\text{open}}$, uncertainty in $\tan\delta$ can be minimized by maximizing $p$ and $Q_\text{open}$.

To assess the impact of measurement uncertainty on loss-tangent extraction --- and how DUT design can mitigate it --- we compare two representative Cases with different participation ratios, following the naming convention used above. Case I corresponds to $p \approx 3\%$, while Case II reaches $p\approx 18\%$, nearly six times higher and close to the maximum achievable participation in this scheme.

In Fig.~\ref{fig3}(b), solid lines show the extracted $\tan\delta~Q_\text{open}$ as a function of $Q_i / Q_\text{open}$ for both Cases, while dashed lines represent the corresponding relative uncertainties of the loss tangent. The fabrication-induced variation is captured by the uncertainty of $Q_\text{open}$, whereas the measurement or circle-fitting uncertainty is represented by the uncertainty of $Q_i$, which is typically orders of magnitude smaller. The figure reveals that the regime for reliable measurements --- where $Q_i$ remains measurable and the extracted loss tangent is accurate --- is inherently narrow and determined by DUT participation and baseline variability.

For low-participation DUTs (Case I), when the DUT loss tangent is comparable to $Q_\text{open}^{-1}$, the resulting change in the total resonator quality factor is minimal, leading to extremely large relative uncertainty in the extracted $\tan\delta$. In contrast, high-participation DUTs (Case II) exhibit at least an order-of-magnitude reduction in uncertainty under the same baseline variability. In both regimes, the uncertainty can be further suppressed by designing DUTs such that the DUT loss dominates over the resonator loss, or by improving $Q_\text{open}$ to minimize the impact of fabrication-induced variation.

Similar in spirit to approaches used in 3D superconducting cavities~\cite{Lei2023}, a multimode self-calibration approach eliminates the need for a separate reference resonator. For each measured mode, Eq.~\ref{eq:resofreq} yields the resonant frequency, which, when inserted into Eq.~\ref{eq:p}, provides the corresponding participation ratio. Together with the measured $Q_i$, Eq.~\ref{eq:Qi_p} then gives one constraint relating $\tan\delta$ and $Q_\text{open}$. Measuring at least two resonant modes yields two equations for these two unknowns, enabling simultaneous extraction of $\tan\delta$ and $Q_\text{open}$ from a single DUT, without reference to a separate resonator.

The above analysis assumes a known frequency dependence of $Q_\text{open}$, which is set by the dominant loss mechanism. \textcolor{black}{In CPW resonators limited by surface or interfacial dielectric loss~\cite{Wenner2011,Bruno2015}, the attenuation constant $\alpha$ increases approximately linearly with frequency~\cite{Gao2008, chen2025}, which should yield a frequency-independent internal quality factor. As demonstrated in Fig.~\ref{fig:Qref_multimode}, our reference resonators exhibit comparable internal quality factors at the fundamental and first harmonic modes, justifying the use of a frequency-dependent $\alpha$ in our simulation.} In contrast, when loss is dominated by weakly conductive dielectrics or resistive elements, $\alpha$ is approximately frequency independent. 
When $Q_\text{open}(f)$ is not known a priori, two modes no longer uniquely determine both $Q_\text{open}$ and $\tan\delta$; access to additional modes is required to constrain the loss mechanism. Conversely, once $Q_\text{open}(f)$ is established, the same formulations naturally enable extraction of $\tan\delta(f)$ across all accessible modes, if such a frequency dependence exists.

\begin{figure*}[t]
\centering
\includegraphics[width=1.75\columnwidth]{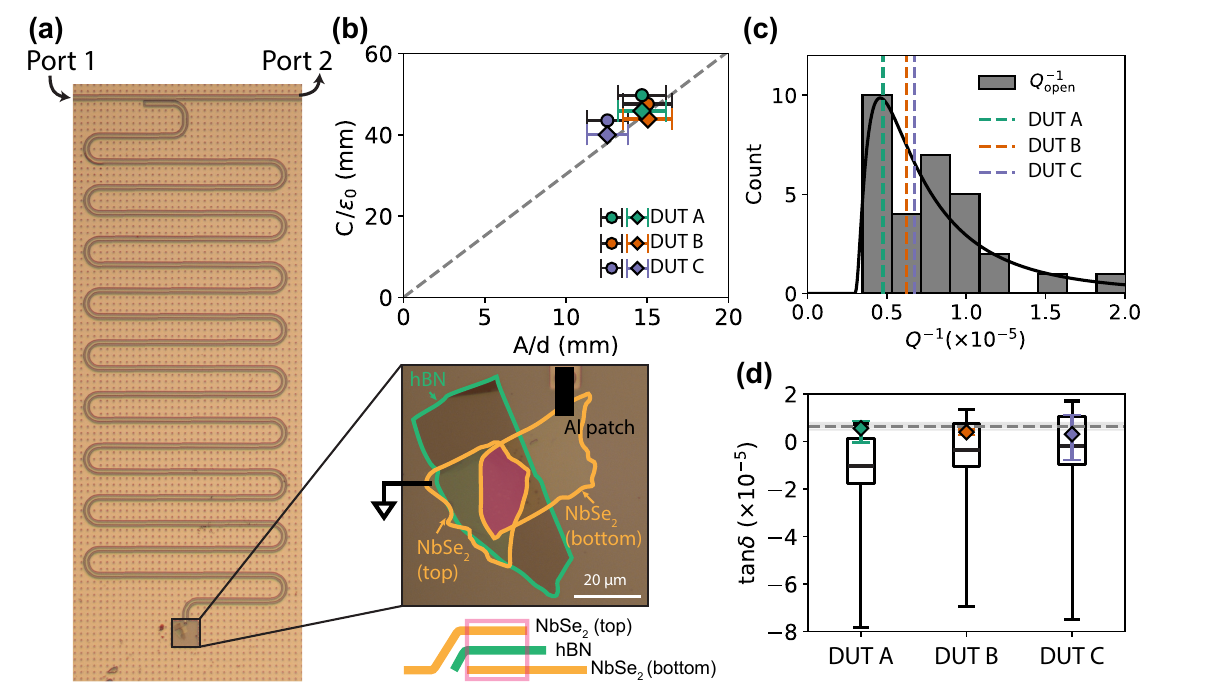}
\caption{
(a) Optical image of a hanger-type Nb CPW resonator terminated by a PPC. The zoomed region highlights the PPC, with colored outlines marking the boundaries of the vdW flakes and the PPC area labeled in pink. An Al patch is included schematically to indicate the electrical connection. The schematic below shows the capacitor cross-section.
(b) Normalized capacitance, $C/\varepsilon_0$, versus the geometric ratio $A/d$ (overlap area $A$ divided by hBN thickness $d$). The gray dashed line represents the PPC model with $\kappa_{\mathrm{hBN}} = 3.03$. Linear fits yields $\kappa_{\mathrm{hBN}} = 3.33 \pm 0.06$ from the single-mode analysis (circles) and $\kappa_{\mathrm{hBN}} = 3.06 \pm 0.08$ after multimode self-calibration (diamonds), demonstrating improved capacitance accuracy after correcting for $f_0$. 
(c) Histogram of inverse quality factor $Q_\text{open}^{-1}$ for reference resonators (gray bars). The data can be described by a log-normal distribution (solid line). Vertical dashed lines indicate the single-photon $Q_i^{-1}$ values at fundamental modes of DUTs A–C.
(d) Extracted loss tangent $\tan\delta$ for DUTs A-C. Box-and-whisker plots show the single-mode results, which rely on comparison to reference resonators. Diamonds with error bars show the multimode results, which eliminate the need for any reference resonator. The gray dashed line and shaded band indicate representative literature values for hBN. The low-$Q$ tail observed in (c) leads to apparent negative outliers in the single-mode $\tan\delta$ extraction; these are removed entirely after multimode calibration, yielding more accurate and statistically tighter results.}
\label{fig4}
\end{figure*}

\section{Experimental Implementation}
To demonstrate the applicability of our explicit closed-form formulations, we present an experimental implementation using fabricated CPW resonators terminated by van der Waals (vdW) parallel-plate capacitors (PPCs) which primarily exhibit capacitive reactance with small, approximately frequency-independent dielectric loss in the microwave regime~\cite{Wang2022}. Although demonstrated with capacitive DUTs, the same formulation (Eqs.~\ref{eq:resofreq},~\ref{eq:p} and~\ref{eq:Qi_p}) applies directly to inductive and hybrid systems. A systematic exploration of these extensions lies beyond the scope of this work but represents a clear direction for future study.

Fig.~\ref{fig4}(a) shows an example of a meandering CPW resonator, all made of niobium, capacitively coupled at its upper end to a horizontal busline. The PPC at the bottom end is created by sandwiching a flake of the vdW dielectric hexagonal boron nitride (hBN) between flakes of the vdW superconductor niobium diselenide (NbSe$_2$). The capacitor area is defined by the overlap of the two NbSe$_2$ flakes, highlighted in pink in the optical image. We measured the transmission coefficient $S_{21}$ through the feedline of three hanger-type resonators terminated by DUT, at a base temperature of 20 mK in a dilution refrigerator. The DUTs were designed to have a high energy participation ratio. 
From circle-fitting of the resonance, we extract $f_r$ and $Q_i$ for both DUT-terminated and open-ended reference resonators. Using these measured quantities and Eqs.~\ref{eq:resofreq} \& \ref{eq:p}, we calculate the capacitance $C_{\text{DUT}}$ and the participation ratio $p$ of the vdW capacitor. From this, we can extract the microwave-frequency dielectric constant $\kappa$ and loss tangent $\tan\delta$ of the hBN. The DUT properties and results are summarized in Table~\ref{tab:data}. For each DUT, both the fundamental ($n=1$) and the second-harmonic ($n=2$) modes are characterized, enabling a comparison between single-mode analysis and the multimode self-calibration approach.

\begin{table*}[htbp]
\caption{\label{tab:data} Properties and measurement results of vdW PPCs}
\begin{ruledtabular}
\begin{tabular}{lcccccccccccc}
 &Area&hBN thickness& & C$_\text{DUT}$ & Reactance & $f_r$ & $f_\text{open}$ & $p$ & $Q_i$
  & \multicolumn{2}{c}{$\tan \delta$ ($\times$10$^{-5}$)} \\
\cline{11-12}

DUT & ({\textmu}m$^2$)\footnotemark[1] & (nm)& Mode & (fF) & ($\Omega$) & (GHz) & (GHz) & \% & ($\times$10$^5$)
      & Median & IQR\\
\hline
A & 300 & 20& $n=1$  & 421 & -120 & 3.410 & 3.941 & 11.43 & 2.103 & -0.37 & [-1.10, 0.79]\\

&  &   & $n=2$  & 388$^\dagger$ & -59 & 6.927 & 3.899$^\dagger$ & 8.12 & 2.215 & 0.56$^\dagger$ & [-0.04, 0.86]$^\dagger$\\

B & 440 & 30& $n=1$ & 440 & -116 & 3.391 & 3.941 & 11.76 & 1.601 & -0.10 & [-0.80, 1.03]\\

&  &   & $n=2$  & 406$^\dagger$ & -57 & 6.900 & 3.897$^\dagger$ & 8.19 & 1.619 & 0.40$^\dagger$ & [0.27,0.51]$^\dagger$\\

C & 439 & 35 & $n=1$ & 385 & -131 & 3.446 & 3.941 & 10.74 & 1.479 & -0.52 & [-1.29, 0.74]\\

&  &   & $n=2$  & 354$^\dagger$ & -64 & 6.980 & 3.900$^\dagger$ & 7.93 & 1.476 & 0.31$^\dagger$ & [-0.77, 1.12]$^\dagger$\\
\end{tabular}
\end{ruledtabular}
\footnotetext[1]{
Area uncertainty $\pm 10\%$.~~~~~~~
$^\dagger$ Obtained from the multimode self-calibration procedure described in the text.}
\end{table*}

Fig.~\ref{fig4}(b) shows the extracted DUT capacitance $C_\text{DUT}$, normalized by the vacuum permittivity $\varepsilon_0$, plotted as a function of the geometric ratio $A/d$, where $A$ is the effective parallel-plate area and $d$ is the hBN thickness. Results obtained from the single-mode analysis are shown as circles, while the multimode self-calibrated results are shown as diamonds.

In a parallel-plate capacitor model, the dielectric constant $\kappa_{\text{hBN}}$ is obtained from the slope of a linear fit to the data with the $y$-intercept constrained to zero. Finite-element simulations were used to confirm that the parallel-plate model is applicable, \textit{i.e.} that fringing fields can be ignored (see Table~\ref{tab:FEM_result}). Frequency errors are negligible compared with geometric uncertainties. The error bars therefore represent uncertainties in the PPC area, estimated to be within 10\% due to imperfections (e.g., bubbles) in the van der Waals interface, which can alter the effective capacitor area. From the fits, we find $\kappa_{\text{hBN}} = 3.33 \pm 0.06$ from the single-mode analysis and $\kappa_{\text{hBN}} = 3.06 \pm 0.08$ after multimode self-calibration of $f_\text{open}$. 
Both values agree well with the reported out-of-plane dielectric constant of bulk hBN, $\kappa_{\text{hBN}} = 3.03$~\cite{Laturia2018}. The closer agreement from multimode calibration arises from the more accurate determination of $f_\text{open}$. We find that the multimode-extracted $f_\text{open}$ is systematically lower than that obtained from the separate open resonator. This discrepancy may result from a small additional inductance associated with the device geometry and/or a slight modification of the effective resonator length introduced by the Al patch. While the exact origin requires further investigation, neglecting this shift leads to an overestimation of $C_{\text{DUT}}$, and therefore an inflated dielectric constant, in a single-mode analysis.








We further compare the hBN loss tangent obtained from the single-mode analysis with that from the multimode self-calibrated approach; the results are summarized in Tab.~\ref{tab:data}. In the single-mode method, the loss tangent is obtained by comparing the single-photon $Q_i$ of DUT-terminated resonators with the statistical distribution of $Q_\text{open}$ measured across nominally identical reference resonators. In contrast, the multimode analysis uses the measured $Q_i$ values from the $n=1$ and $n=2$ modes of the same DUT to directly solve for both $\tan\delta$ and $Q_\text{open}$, again noting that $Q_\text{open}$ is independent of frequency. 

Fig.~\ref{fig4}(c) shows the distribution of $Q_\text{open}^{-1}$ for the reference resonators, together with the fundamental-mode $Q_i^{-1}$ values of DUTs A–C (vertical dashed lines). The reference resonators exhibit a broad spread in $Q_\text{open}$, well described by a log-normal distribution \cite{McRae2025}. Because hBN exhibits low dielectric loss, the DUT $Q_i$ values lie near the center of this distribution (around $1.48\times10^5$), where the DUT introduces only a small change to the total resonator loss. As a result, the inferred $\tan\delta$ becomes highly sensitive to device-to-device variability in $Q_\text{open}$ rather than to the intrinsic loss of the hBN.

To quantify this effect, we extract $\tan\delta$ using Eq.~\ref{eq:Qi_p} by pairing each measured DUT $Q_i$ with all sampled values of $Q_\text{open}$, generating a distribution of inferred loss tangent values shown in Fig.~\ref{fig4}(d). The median values for DUTs A–C are $(-1.04, -0.37, -0.20)\times10^{-5}$ with interquartile ranges of approximately $\pm1\times10^{-5}$. The negative medians arise from sampling the low-$Q$ tail of the reference distribution and do not indicate physical negative loss. This behavior demonstrates that single-mode extraction becomes unreliable for low-loss materials when $Q_\text{open}$ varies significantly across nominally identical resonators.

The multimode analysis overcomes this limitation by directly solving for $\tan\delta$ and $Q_\text{open}$ using two resonant modes of the same DUT. Using Eqs.~\ref{eq:resofreq}, \ref{eq:p}, and \ref{eq:Qi_p}, we extract the loss tangent. The resulting values for DUTs A–C are $(5.57, 3.97, 3.11)\times10^{-6}$, shown as diamonds with error bars in Fig.~\ref{fig4}(d). In this approach, no comparison to reference resonators is required; the error bars arise solely from the spread in the measured $Q_i$ across the available modes (see Fig.~\ref{fig:Qi_data_multimode}), leading to substantially reduced uncertainty and improved  internal consistency compared with the single-mode distributions. The extracted hBN loss tangent agrees well with previously reported upper bounds of $(5.0$–$7.8)\times10^{-6}$~\cite{Wang2022}, shown as the shaded band.


Taken together, these results validate the applicability of our analytic formulations and establish multimode self-calibration as a robust, reference-free method for extracting circuit parameters and dielectric loss. Future improvements in resonator design and measurement bandwidth will allow access to additional modes within the same DUT, enabling systematic studies of mode-dependent reactance and loss mechanisms in quantum materials.

\section{Conclusion}
We present a general analytic framework for extracting reactive circuit parameters and loss tangent in transmission-line resonators, based on Eqs.~\ref{eq:resofreq}, \ref{eq:p}, and \ref{eq:Qi_p}. Unlike prior microwave analyses that rely on perturbative approximations or computationally intensive electromagnetic simulations, this framework applies to arbitrary reactive loads and enables efficient, simulation-free microwave characterization of nanoscale and low-dimensional materials.

The analysis further identifies the regime of maximal extraction accuracy, which occurs when the DUT reactance is comparable to the resonator impedance and the DUT energy participation ratio is maximized. We validate the framework through circuit simulations and low-temperature measurements of van der Waals capacitors, demonstrating both quantitative accuracy and experimental feasibility. In addition, the multimode self-calibration approach mitigates uncertainties arising from reference-resonator variability by solving for both $f_\text{open}$ and $Q_\text{open}$ within a single device. Beyond improving accuracy, this multimode analysis provides physical discrimination between capacitive and inductive components of the DUT reactance. \textcolor{black}{Once the effective DUT reactance is experimentally determined, electromagnetic simulations can be used, if desired, to relate it to specific device geometry and material properties.}

Together, these results establish a unified resonator-based metrology framework that connects circuit theory with experiment and enables rapid extraction of capacitance, inductance, and loss tangent. The method is broadly applicable to microwave circuits and material interfaces, and is particularly suited for quantum circuit platforms where precise device characterization is essential.

\textbf{ACKNOWLEDGEMENTS} This work was primarily supported by the Army Research Office under Contract W911NF-22-C-0021 (NextNEQST SuperVan-2). Synthesis of boron nitride (K.W.~and T.T.) was supported by the Elemental Strategy Initiative conducted by the MEXT, Japan (Grant No.~JPMXP0112101001) and JSPS KAKENHI (Grant Nos.~JP19H05790 and JP20H00354). J.H.~acknowledges support from the Gordon and Betty Moore Foundation’s EPiQS Initiative, Grant GBMF10277. J.P.~acknowledges support from the National Research Foundation of Korea(NRF) grant funded by the Korea government(MSIT)(RS-2024-00358841, RS-2024-00393599, IITP-2025-RS-2022-00164799) and the education and training program of the Quantum Information Research Support Center, funded through the NRF by the MSIT of the Korean government (No.~2021M3H3A1036573). J.B.~acknowledges support from the Army Research Office under Grant Number W911NF-24-1-0133. The views and conclusions contained in this document are those of the authors and should not be interpreted as representing the official policies, either expressed or implied, of the Army Research Office or the U.S. Government. The U.S. Government is authorized to reproduce and distribute reprints for Government purposes notwithstanding any copyright notation herein.


\appendix

\section{Resonant Condition}
\label{appendix:f_derivation}

Using the telegrapher’s equations, the voltage and current at position $z$ along the resonator are given by
\begin{align}
    V(z) &= V_0^+ e^{-\gamma z} + V_0^- e^{\gamma z}, \label{eq:V} \\
    I(z) &= \frac{V_0^+}{Z_0} e^{-\gamma z} - \frac{V_0^-}{Z_0} e^{\gamma z}, 
\end{align}
where $V_0^+$ and $V_0^-$ are the forward- and backward-propagating voltage amplitudes at $z=0$.  
The propagation constant is defined as $\gamma = \alpha + j\beta$, where $\alpha$ is the attenuation constant per unit length, $j$ is the imaginary unit, and $\beta = \omega\sqrt{LC}$ is the phase constant. Here, $L$ and $C$ are the inductance and capacitance per unit length of the resonator, respectively. The characteristic impedance is $Z_0 = \sqrt{L/C}$.  

The phase constant $\beta$ depends on frequency such that $\beta \ell = \pi f_r / f_\text{open}$, where $f_\text{open} = 1 / (2\ell \sqrt{LC})$ is the fundamental resonance frequency of the open-ended resonator. In the main text, we introduced a dimensionless phase parameter $\phi = 2\pi {f_r}/{f_\text{open}}$ to quantify the phase shift imposed by the DUT termination. This parameter is directly related to the phase constant through the expression
$\phi = 2\beta\ell.$
We define the load impedance as $Z_L = r + jX$, where $r$ and $X$ denote the resistive and reactive components, respectively, and the loss tangent is $\tan\delta = r / \lvert X\rvert $.

When the coupling capacitance is sufficiently small, a voltage antinode forms at $z = -\ell$ at resonance, i.e., $\lvert V(z=-\ell)\rvert $ reaches its maximum value. Evaluating Eq.~\eqref{eq:V} at $z = -\ell$ gives
\begin{align}
    V(z=-\ell) = V_0^+ e^{j\gamma \ell} \bigl(1 + \lvert\Gamma\rvert  e^{-2\alpha \ell} e^{j(\theta - 2\beta \ell)} \bigr), \label{eq:Vzl}
\end{align}
where $\Gamma$ is the reflection coefficient, defined as
\begin{align}
    \Gamma &= \lvert\Gamma\rvert  e^{j\theta} = \frac{V_0^-}{V_0^+} = \frac{Z_L - Z_0}{Z_L + Z_0}.
\end{align}
When $r\ll<X$, the phase $\theta$ and magnitude $\lvert\Gamma\rvert $ are approximately
\begin{align}
    \theta &\approx \tan^{-1}\!\left( \frac{2 X Z_0}{X^2 - Z_0^2} \right), \label{eq:theta}\\
    \lvert\Gamma\rvert  &\approx 1 - \frac{2 Z_0 r}{X^2 + Z_0^2} \approx 1 - \sin\theta \, \tan\delta.
\end{align}

From Eq.~\eqref{eq:Vzl}, $\lvert V(z=-\ell)\rvert $ is maximized when $e^{j(\theta - 2\beta \ell)} = 1$, leading to the resonance condition
\begin{align}
    \theta - 2\beta \ell = 2n\pi. \label{eq:ResCon}
\end{align}
Thus, the resonant frequency $f_r$ of the circuit is
\begin{align}
    \frac{f_r}{f_\text{open}} = \frac{1}{\pi} \tan^{-1}\!\left( \frac{Z_0}{X} \right) + n,
\end{align}
where we have used the trigonometric identity $\tan(\theta / 2) = Z_0 / X$ from Eq.~\eqref{eq:theta}.

\section{Energy participation ratios}
\label{appendix:p_derivation}

By definition, the participation ratio is
\begin{align}
 p &= \frac{\langle W_\text{DUT} \rangle}{\langle W_\text{tot} \rangle},
\end{align}
where $\langle W \rangle$ denotes the time-averaged stored electromagnetic energy, and $\langle W_\text{tot} \rangle = \langle W_\text{res} \rangle + \langle W_\text{DUT} \rangle$.

We consider a capacitive DUT, for which the stored energy is purely electric. The resonator stores both electric and magnetic energy, denoted by $\langle W_\text{res,e} \rangle$ and $\langle W_\text{res,m} \rangle$, respectively:
\begin{align}
\langle W_\text{res,e}\rangle &= \int_{-\ell}^{0} dz \, \frac{1}{4} C \lvert V(z)\rvert ^2 \\
&\approx \frac{\lvert V_0^+\rvert ^2 (1 + \lvert\Gamma\rvert ^2)}{8 f_\text{open} Z_0} \left[ 1 + \frac{\sin \phi}{\phi} \right], \label{eq:W_res_e}\\
\langle W_\text{res,m}\rangle &= \int_{-\ell}^{0} dz \, \frac{1}{4} L \lvert I(z)\rvert ^2 \\
&\approx \frac{\lvert V_0^+\rvert ^2 (1 + \lvert\Gamma\rvert ^2)}{8 f_\text{open} Z_0} \left[ 1 - \frac{\sin \phi}{\phi} \right]. \label{eq:W_res_m}
\end{align}

The energy stored in the capacitive DUT is
\begin{align}
\langle W_\text{DUT,C}\rangle &= \frac{1}{4} C_\text{DUT} \lvert V(z{=}0)\rvert ^2 \lvert \frac{X}{Z_L} \rvert^2 \\
&\approx \frac{\lvert V_0^+\rvert ^2 (1 + \lvert\Gamma\rvert ^2)}{4 f_\text{open} Z_0} \lvert \frac{\sin \phi}{\phi} \rvert, \label{eq:energy_in_device}
\end{align}
with $\sin \phi < 0$.

At resonance, the time-averaged electric and magnetic energies are balanced such that
\[
\langle W_\text{res,e}\rangle + \langle W_\text{DUT,C}\rangle = \langle W_\text{res,m}\rangle.
\]
For an inductive DUT, the same derivation applies, and Eqs.~\eqref{eq:W_res_e}–\eqref{eq:energy_in_device} remain valid with $\sin \phi > 0$, while the energy balance becomes
\[
\langle W_\text{res,m}\rangle + \langle W_\text{DUT,L}\rangle = \langle W_\text{res,e}\rangle.
\]

Combining Eqs.~\eqref{eq:W_res_e}–\eqref{eq:energy_in_device}, the participation ratio can be written as
\begin{align}
p &= \frac{\lvert\sin \phi\rvert}{\phi + \lvert\sin \phi\rvert}.
\end{align}

\section{Q factors}
\label{appendix:q_derivation}

The internal quality factor $Q_i$ of the resonator--DUT system can be written as
\begin{align}
Q_i^{-1} = Q_\text{res}^{-1} + Q_\text{DUT}^{-1}, \label{Q_i}
\end{align}
where $Q_\text{res}$ and $Q_\text{DUT}$ denote the quality factors of the resonator and the DUT, respectively, evaluated at the resonant frequency $f_r$.

These are defined by
\begin{align}
Q_\text{res}^{-1} &= \frac{1}{2\pi f_r} \frac{P_\text{res}}{\langle W_\text{tot} \rangle}, \\
Q_\text{DUT}^{-1} &= \frac{1}{2\pi f_r} \frac{P_\text{DUT}}{\langle W_\text{tot} \rangle},
\end{align}
where $P_\text{res}$ and $P_\text{DUT}$ represent the power dissipated in the resonator and in the DUT.

The power dissipated in the DUT is
\begin{align}
P_\text{DUT} &= \frac{1}{2} \mathrm{Re}\left[V(0) I^*(0)\right] \\
&= \frac{\lvert V_0^+\rvert ^2}{2 Z_0} \left(1 - \lvert\Gamma\rvert ^2\right),
\end{align}
while the power loss in the resonator is given by
\begin{align}
P_\text{res} &= \frac{1}{2} \mathrm{Re}[V(-\ell) I^*(-\ell)] - p \\
&\approx \frac{\lvert V_0^+\rvert^2}{Z_0} \alpha \ell (1 + \lvert\Gamma\rvert^2).
\end{align}

These lead to the following expressions for the inverse quality factors:
\begin{align}
Q_\text{DUT}^{-1} &= \frac{2 \lvert\sin \phi\rvert}{\phi + \lvert\sin \phi\rvert} \tan \delta \\
&\approx 2p \tan \delta, \\
Q_\text{res}^{-1} &\approx \frac{4 \alpha \ell}{\phi + \lvert\sin \phi\rvert} \\
&= (1 - p) \frac{2\pi}{\phi} Q_\text{open}^{-1},
\end{align}
where $Q_\text{open}^{-1} = 2 \alpha \ell / \pi$ is the internal quality factor of an equivalent open-ended transmission-line resonator, consistent with the definition used in the main text.

Substituting the above expressions into Eq.~\eqref{Q_i}, the internal quality factor becomes
\begin{align}
Q_i^{-1} = \frac{2 \lvert\sin \phi\rvert \tan \delta + 2\pi Q_\text{open}^{-1}}{\phi + \lvert\sin \phi\rvert},
\end{align}
which can also be expressed in terms of frequency if desired.

\section{Device fabrication}
\label{fab}
The coplanar waveguide (CPW) resonators are fabricated from niobium (Nb) films deposited on high-resistivity Si substrates. A 200-nm-thick Nb layer is sputtered onto the Si wafer, followed by photolithographic patterning and dry etching to define the resonator geometry. Prior to transferring the vdW capacitor stack, the Nb chip is cleaned sequentially using AR-600 resist remover, a piranha solution (H$_2$SO$_4$ : H$_2$O$_2$ = 3:1) to remove organic residues, and a 10:1 buffered oxide etchant (BOE) to eliminate the excess oxide formed during the piranha process. The chip is then immediately transferred into a nitrogen glovebox (H$_2$O, O$_2$~<~0.5~ppm) to minimize reoxidation and stored there until stacking.

The vdW capacitor is assembled entirely inside the glovebox under an inert environment to prevent degradation of air-sensitive materials. Both NbSe$_2$ and hBN flakes are freshly exfoliated before assembly: NbSe$_2$ is exfoliated inside the glovebox using PDMS and blue tape, while hBN is exfoliated in air using Scotch tape. We employ a dry transfer technique using a polycarbonate (PC) stamp to pick up and stack the layers, followed by melting and dropping off of the completed stack onto the pre-patterned Nb resonator chip. The PC film is subsequently dissolved in chloroform.

To define the electrical contacts, we spin-coat the chip with a bilayer of PMMA A4/MMA EL10 resist and pattern the contact regions using electron-beam lithography. These contacts connect one side of the NbSe$_2$ flake to the resonator center conductor and the other side to the ground plane. Before metal deposition, we perform \emph{in situ} Ar ion milling to remove surface oxides from both the NbSe$_2$ and Nb contact regions. Finally, aluminum (Al) is deposited to form high-transparency superconducting contact between the vdW material and the Nb resonator \cite{Antony2021}.

\section{Supplementary Information}
\label{SI}

Figures and tables providing additional supporting data are summarized as follows:
\begin{enumerate}
    \item Schematic of the experimental setup.
    \item Extracted single-photon $Q_i$ values for DUTs~A–C in different resonant modes.
    \item Reference resonator $Q_{i,n}$ at multiple modes ($n=1,2$) versus photon number.
    \item Parameters of reference resonators at the single-photon limit.
    \item Finite-element simulation results.
\end{enumerate}

\begin{figure}[h]
\includegraphics[width=1\columnwidth]{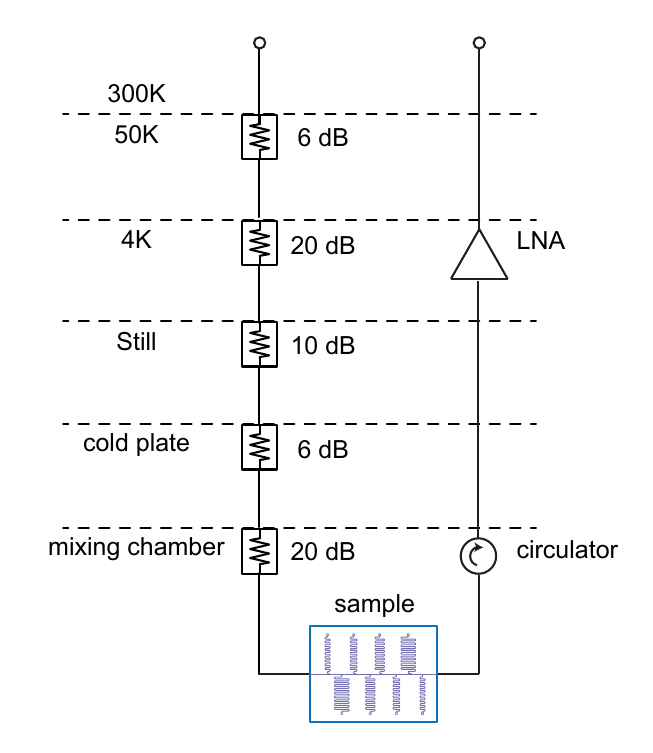}
\caption{Experimental setup used for resonator characterization. The total attenuation of the input microwave line from room temperature to the mixing chamber is approximately 65~dB. A vector network analyzer (VNA) is used as both the microwave source and detector.}
\label{fig:setup}
\end{figure}

\begin{figure}[h]
\includegraphics[width=1\columnwidth]{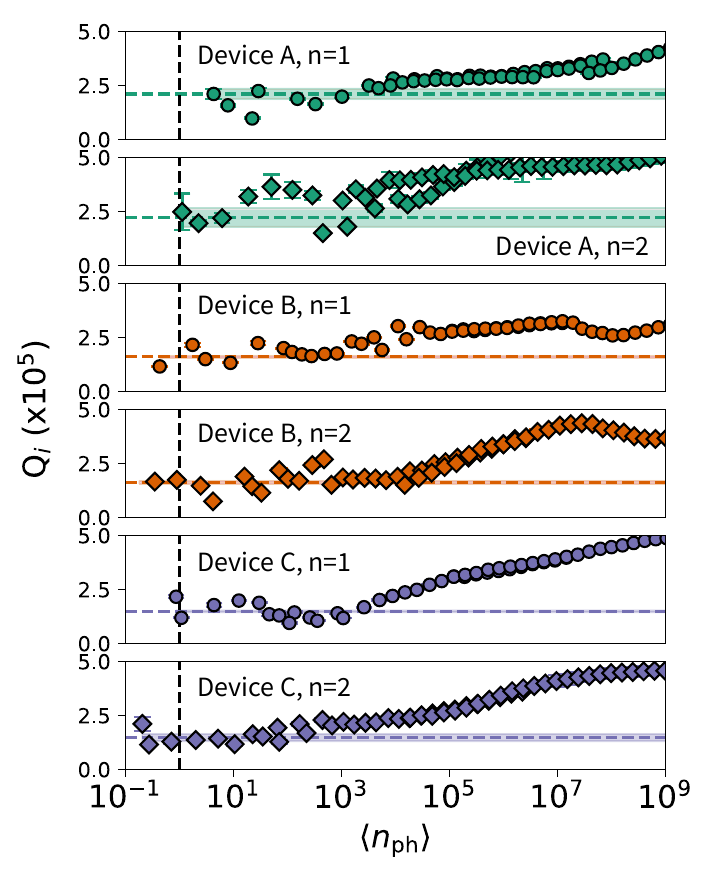}
\caption{Internal quality factor $Q_i$ as a function of the mean photon number $\langle n_{\mathrm{ph}}\rangle$ for DUTs~A–C, vertically offset for clarity. 
Data from the fundamental mode are shown as circles, while those from the second harmonic ($n=2$) mode are shown as diamonds. 
The single-photon $Q_i$ values are indicated by horizontal dashed lines, with shaded bands representing the associated uncertainties. 
The numerical values corresponding to the single-photon points are listed in Table~\ref{tab:data}.}\label{fig:Qi_data_multimode}\end{figure}

\begin{figure}[h]
\includegraphics[width=1\columnwidth]{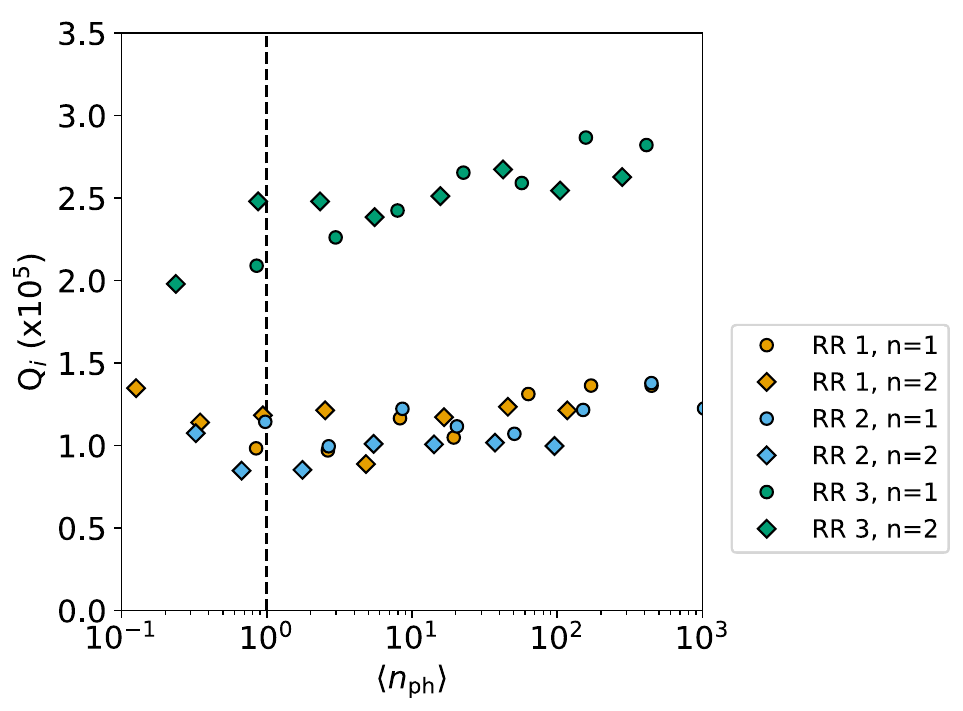}
\caption{Internal quality factor $Q_i$ as a function of the mean photon number for reference resonators (RR) ~1--3. The vertical dashed line represents the single-photon limit. The reference resonators were fabricated using the same process steps—including ion milling and Al patching at the termination end—as those used for the DUT-integrated resonators. The extracted frequency and quality factors at the single-photon limit for the fundamental and second harmonic modes are summarized in Table~\ref{tab:q_ref_in_appendix}. The results show that the reference resonators exhibit nearly frequency-independent quality factors across different modes.}
\label{fig:Qref_multimode}
\end{figure}

\begin{table}[htbp]
\caption{\label{tab:q_ref_in_appendix}Parameters of reference resonators at the single-photon limit.}
\begin{ruledtabular}
\begin{tabular}{lccccc}
 &  & $f_{n=1}$ & $f_{n=2}$ & $Q_{i,n=1}$ & $Q_{i,n=2}$ \\
Resonator & Type & (GHz) & (GHz) & $(\times10^5)$ & $(\times10^5)$ \\
\hline
RR~1 & short-ended & 2.971 & 8.865 & 0.983 & 1.183 \\
RR~2 & short-ended & 2.980 & 8.895 & 1.143 & 0.847 \\
RR~3 & open-ended  & 3.941 & 7.888 & 2.089 & 2.480 \\
\end{tabular}
\end{ruledtabular}
\end{table}

\begin{table}[htbp]
\caption{\label{tab:FEM_result}Comparison of capacitance values obtained from analytic calculation, finite-element simulation, and experimental measurement. Finite-element simulations were performed using \textit{Ansys Electronics Desktop}. The analytic capacitance was estimated using the parallel-plate capacitor model with corrections for edge effects.}
\begin{ruledtabular}
\begin{tabular}{lccc}
DUT & $C_{\mathrm{analytic}}$ (fF) & $C_{\mathrm{sim}}$ (fF) & $C_{\mathrm{meas}}$ (fF) \\
\hline
A & 399 & 400 & 388 \\
B & 367 & 370 & 406 \\
C & 336 & 339 & 354 \\
\end{tabular}
\end{ruledtabular}
\end{table}

\clearpage
\bibliographystyle{apsrev4-2}
\bibliography{reference}

\end{document}